%
\input harvmac
\def\Title#1#2#3{#3\hfill\break \vskip -0.35in
\rightline{#1}\ifx\answ\bigans\nopagenumbers\pageno0\vskip.2in
\else\pageno1\vskip.2in\fi \centerline{\titlefont #2}\vskip .1in}


\def\bra#1{\langle #1 |}
\def\ket#1{| #1\rangle}
\def\braket#1#2{\langle \, #1 \, | \, #2 \, \rangle}

\def\S{{\cal{S}}}

\def\H{{\cal{H}}}

\def\R{\hbox{\rm I \kern-5pt R}}

\font\ticp=cmcsc10
\def\ajou#1&#2(#3){\ \sl#1\bf#2\rm(19#3)}
%
%
\lref\griff{R.B.~Griffiths, \ajou J. Stat. Phys. &36 (84) 219.}

\lref\grifflogic{R.B.~Griffiths, \ajou Found. Phys. &23 (93) 1601.}

\lref\griffchqr{R.B.~Griffiths, \ajou Phys. Rev. A &54 (96) 2759.} 

\lref\ishamlinden{C.J.~Isham, \ajou J. Math. Phys. &23 (94) 2157; 
C.J.~Isham and N.~Linden, \ajou J. Math. Phys. &35 (94) 5452.}

\lref\omnes{R.~Omn\`es, \ajou J. Stat. Phys. &53 (88) 893.}

\lref\omnesbook{R.~Omn\`es, {\it The Interpretation of Quantum
Mechanics}, Princeton University Press, Princeton (1994).}

\lref\omnesquote{Page 161 of \omnesbook .} 

\lref\omnesfp{R.~Omn\`es, \ajou Found. Phys. &25 (95) 605.} 

\lref\dkretro{F.~Dowker and A.~Kent, unpublished.}

\lref\gmhsantafe{M.~Gell-Mann and J.B.~Hartle in {\it Complexity, Entropy,
and the Physics of Information, SFI Studies in the Sciences of
Complexity}, Vol.
VIII, W.~Zurek (ed.),  Addison Wesley, Reading (1990).}

\lref\gmhthree{M.~Gell-Mann and J.B.~Hartle in {\it Proceedings of the
NATO Workshop on the Physical Origins of Time Asymmetry, Mazag\'on, Spain,
September 30-October 4, 1991}, J.~Halliwell, J.~P\'erez-Mercader, and
W.~Zurek (eds.), Cambridge University Press, Cambridge (1994); gr-qc/9304023.}

\lref\gmhprd{M.~Gell-Mann and J.B.~Hartle, \ajou Phys. Rev. D
&47 (93) 3345.}

\lref\gmhequiv{M.~Gell-Mann and J.B.~Hartle, gr-qc/9404013 (unpublished).}

\lref\akordered{A.~Kent, ``Quantum Histories and Their Implications'',
gr-qc/9607073, submitted to Ann. Phys.}

\lref\dowkerkentone{F.~Dowker and A.~Kent, 
\ajou J. Stat. Phys. &82 (96) 1575.}

\lref\dowkerkenttwo{F.~Dowker and A.~Kent, \ajou Phys. Rev. Lett. &75
(95) 3038.}

\lref\kenttwo{A.~Kent, 
\ajou Phys. Rev. A &54 (96) 4670.}

\lref\abl{Y.~Aharonov, P.~Bergmann and J.~Lebowitz, 
\ajou Phys. Rev. B &134 (64) 1410.} 

\lref\cohen{O.~Cohen, \ajou Phys. Rev. A &51 (95) 4373.}

\lref\av{Y.~Aharonov and L.~Vaidman, \ajou J. Phys. A &24 (91) 2315.} 

\Title{
\vbox{\baselineskip12pt\hbox{ DAMTP/96-18}\hbox{ gr-qc/9604012}{}}
}
{\vbox{\centerline {Consistent Sets Yield Contrary Inferences in
Quantum Theory}}}{~}

\centerline{{\ticp Adrian Kent\foot{Email: apak@damtp.cam.ac.uk}}}
\vskip.1in
\centerline{\sl Department of Applied Mathematics and
Theoretical Physics,}
\centerline{\sl University of Cambridge,}
\centerline{\sl Silver Street, Cambridge CB3 9EW, U.K.}

\bigskip

\centerline{\bf Abstract}
{
In the consistent histories formulation of quantum theory,
the probabilistic predictions and retrodictions made from 
observed data depend on the choice of a consistent set.  
We show that this 
freedom allows the formalism to retrodict
contrary propositions which correspond to orthogonal commuting 
projections and which each have probability one.
We also show that the formalism makes contrary 
probability one predictions when applied 
to Gell-Mann and Hartle's generalised time-neutral 
quantum mechanics. 
\vfill
To appear in Phys. Rev. Lett.}
\eject
\newsec{Introduction} 
The consistent histories approach to quantum theory, pioneered by 
Griffiths\refs{\griff, \grifflogic}, Omn\`es\refs{\omnes, \omnesbook},
and  Gell-Mann and Hartle\refs{\gmhsantafe, \gmhthree, \gmhprd}, 
is perhaps the best attempt to date at a precise formulation of 
quantum theory that involves no ``hidden'' auxiliary variables
and can be applied to closed systems.
Since modern ideas in cosmology and quantum gravity require
an interpretation of the quantum theory of the universe,  
the approach has naturally attracted a good deal of interest.
It seems, however, that without 
new axioms, whose precise form is presently 
unknown, it is impossible to derive
the standard predictions of classical mechanics and Copenhagen 
mechanics from the consistent histories formalism, since 
predictions essentially always depend not only on known data and
the hamiltonian but on the selection of a particular
consistent set\refs{\dowkerkentone}. 
This is true even if predictions are made conditional on the
persistence of quasiclassical physics\refs{\kenttwo}.  
It should be said, nonetheless, that debate over the 
scientific status of the formalism continues and that
its proponents tend to regard its lack of predictive power with 
more equanimity than do its critics.

This letter, though, looks at the logical properties
of the consistent histories formalism rather than interpretational
questions.  It is helpful immediately to introduce a little
logical terminology.  
We say that two projection operators $P$ and $Q$ are {\it complementary}
if they fail to commute: $PQ \neq QP$. 
We say that they are {\it contradictory} if they sum to the identity, 
so that $ P = 1 - Q$ and $PQ = QP = 0$. 
Finally, we say that they are {\it contrary} if they are orthogonal
and not contradictory, so that $ P < 1 - Q$ and again $PQ = QP = 0$. 

The somewhat counterintuitive 
properties of consistent sets of histories 
have, of course, already been extensively 
investigated in the original literature\refs{\griff, \omnes,
\gmhsantafe} and elsewhere\refs{\dowkerkentone, \dowkerkenttwo}. 
We describe here a feature which seems to have gone unnoticed, 
namely that different consistent sets extending a given history can 
imply, with probability one, propositions which are 
contrary. 

It is well known that the predictions and retrodictions
made in different sets generally correspond to complementary 
projections and so are incompatible.  This, in itself, might 
be taken as merely an extension to the consistent histories
context of the familiar fact that non-commuting observables 
cannot simultaneously be assigned values.   
The inferences we consider here, though, correspond to 
commuting but orthogonal projections.  
The fact that they are nonetheless each assigned probability
one in their respective sets
is a result with no parallel in standard quantum theory. 
It raises the question of 
whether the present version of the consistent histories formalism
is a natural generalisation of Copenhagen quantum mechanics. 

\newsec{The Consistent Histories Formalism}

We begin with a brief description of the simplest version 
of the consistent histories 
formulation of non-relativistic quantum mechanics, in which
sets of histories correspond to sets of projective decompositions.
While consistent histories can be defined abstractly on any
Hilbert space $\cal{H}$, it is generally assumed
that operators corresponding to 
the hamiltonian $H$ and other physically interesting observables, 
such as position, momentum and spin, are given. 
The dynamics, however, are irrelevant to the 
examples we consider, so that we will take $H=0$ and will not 
need to distinguish any particular operators as 
simple physical observables.  

We are interested in a closed system whose initial density 
matrix $\rho$ is given.  
We will also be interested in applying the formalism to a
version of the time-neutral generalisation of quantum mechanics
first discussed by Aharonov, Bergmann and Lebowitz\refs{\abl}. 
This can be thought of as a 
non-relativistic version of the theory which would be obtained 
by imposing initial and final conditions in 
quantum cosmology, and requires an initial density
matrix, which we take to have the standard normalisation and 
write as $\rho_i$, and final boundary conditions defined by a 
positive semi-definite matrix
$\rho_f$ normalised so that $\Tr ( \rho_i \rho_f ) = 1$.
The initial and final matrices $\rho_i$ and $\rho_f$ then give
boundary conditions for the system at times $t_i$ and $t_f$,
with $t_i < t_f$.  

It may be helpful to think of the system under discussion
as a non-self-interacting quantum system that is prepared 
in the state $\rho_i$ at $t=t_i$ and then 
isolated until $t=t_f$, when it is observed in the
state $\rho_f$.  The Hilbert space $\cal{H}$ here should then
be thought of as a subspace of the full Hilbert space, which
includes the degrees of freedom of the preparation and measuring 
apparatuses, and the full Hamiltonian is nonzero.  
In the examples we consider, both $\rho_i$ and $\rho_f$ are pure 
and $\cal{H}$ is three-dimensional or higher.  Such examples can, 
of course, easily be realised in the laboratory.   

The physical propositions we are interested in
correspond to members of sets $\sigma$ of orthogonal 
hermitian projections $P^{i}$, with
\eqn\conds{\sum_i P^{i} = 1 \quad {\rm{and}}\quad
P^{i} P^{j} = \delta_{ij} P^{i}.}
These projective decompositions of the identity should be 
thought of as being applied at definite times.
The times are usually appended to the sets
of projections, so that 
$\sigma_j (t_j) =\{ P^{i}_j (t_j ) ; i = 1,2,\ldots , n_j \}$ 
defines a set of projections obeying \conds\ and applied at time
$t_j$.
However, as our results depend only on the time ordering, 
we will omit explicit time labels and take sets 
of the form $\S = \{ \sigma_1 , \ldots , \sigma_n \}$ to 
be ordered with time increasing from left to right. 
The projections correspond to propositions about 
the system in the usual way.  For example, a projection onto
the $\sigma_z = 1/2$ eigenspace of a spin-$1/2$ particle applied at
time $t$ corresponds to the statement that the particle 
was in the $\sigma_z = 1/2$ eigenstate at the relevant time. 
The consistent histories formulation differs
from Copenhagen quantum mechanics, however, in that there
is no dynamical projection postulate attached to statements of
this type.

Suppose now we have a list of sets $\sigma_j $ of this form.  
Then the histories given by choosing one projection from each
$\sigma_j$ in all possible ways are an exhaustive and exclusive set of
alternatives. 
We use Gell-Mann and Hartle's decoherence condition, and
say that $\S$ is a {\it consistent} set of histories if
\eqn\decohgmha{ \Tr (  P^{i_n}_n \ldots P^{i_1}_1
                               \rho P^{j_1}_1
\ldots P^{j_n}_n ) =
\delta_{i_1 j_1 } \ldots \delta_{i_n j_n } p(i_1 \ldots i_n ) \, ,}
or, in the time-neutral case, 
\eqn\decohgmh{ \Tr ( \rho_f P^{i_n}_n \ldots P^{i_1}_1
                               \rho_i P^{j_1}_1
\ldots P^{j_n}_n ) =
\delta_{i_1 j_1 } \ldots \delta_{i_n j_n } p(i_1 \ldots i_n ) \, .}
When $\S$ is consistent, $p(i_1 \ldots i_n )$ is the probability of the
history $\{ P^{i_1}_1 , \ldots , P^{i_n}_n \}$. 
We say the set
\eqn\extend{
\S' =  \{\sigma_1,\dots, \sigma_k, \tau, \sigma_{k+1},\dots,
\sigma_n\}}
is a {\it consistent extension} of a consistent set of histories
$\S =  \{\sigma_1 , \ldots , \sigma_n \}$ by 
the set of projections $\tau = \{Q^i : i = 1, \dots, m\}$ if $\tau$
is a projective decomposition and $\S'$ is consistent.  

Suppose now that we have a collection of data defined by the history 
\eqn\history{H = \{ P^{i_1}_1 , \ldots , P^{i_n}_n \}}
which has non-zero
probability and belongs to the consistent set $\S$. 
This history might, for example, describe the results of a series
of experiments or the observations made by an
observer.  To make scientific use of the formalism we 
then want to make further inferences from the data. 
In the standard formalism, this can only be done relative to a 
choice of consistent extension $\S'$ of $\S$.\foot{
We consider only the standard formalism here.  It is
possible to amend the formalism by appending axioms which 
identify natural retrodictions\refs{\omnes, \omnesfp, \dkretro}. 
If only these retrodictions are allowed, contrary retrodictions are avoided.
However, this would also exclude almost all scientifically
desirable retrodictions\refs{\dkretro}.}
Once $\S'$ is fixed we can make probabilistic 
inferences conditioned on the history $H$. 
For example, if $\S'$ has the above form, the histories
extending $H$ in $\S'$ are
$H^i = \{ P^{i_1}_1 , \ldots , 
P^{i_k}_k , Q^i , P^{i_{k+1}}_{k+1}, \ldots , P^{i_n}_n \}$
and the history $H^i$ has conditional probability $p( H^i ) / p(H)$.

We use the convention
that the calculation is made at the time of the last event from
the history, so that any projection occurring before this last
event is a retrodiction.
Thus if $k=n$ then $p(H^i )/p(H)$ is the 
probability with which the proposition 
corresponding to the projection $Q^i$ is predicted; 
if $k<n$ it is the probability with which the proposition is
retrodicted.   
The different $\S'$ are to be thought of as 
different, equally valid, possible pictures of the past and future 
physics of the system, or more formally as different and generally 
incompatible logical structures allowing different classes of inferences 
from the given data.  

\newsec{Contrary retrodictions and predictions} 

We now give two simple examples of contrary inference
in the consistent histories formalism. 
The Hilbert space $\cal{H}$ is taken to be
of dimension greater than or equal to three. 
\medskip\noindent
{\it Example 1}  \qquad  Take $\rho =  \ket{a} \bra{a}$, where the 
normalised vector $\ket{a}$ defines the initial state of the system.  
Define the
projection $P_c = \ket{ c} \bra{c}$, for some 
normalised vector $c$ such that $0 < \, | \braket{a}{c} | \, \leq 1/3$. 
Suppose that the data correspond to the history 
$\H = \{ P_c \}$ from the 
consistent set $\S = \{ \{P_c , 1 - P_c \} \}$. 
Now consider a consistent extension of the form 
$\S' = \{ \{P_b , 1 - P_b \}, \{P_c , 1 - P_c \} \}$, 
where $P_b = \ket{b} \bra{b} / {| \braket{b}{b} |}$ for some 
unnormalised vector $\ket{b}$ with the property
that 
\eqn\vone{ \braket{c}{b} \, \braket{b}{a} \, = \, 
\braket{c}{a} \, \braket{b}{b} \, . } 
It is not hard to verify that $\S'$ is consistent and that 
the conditional probability of $P_b$ given $H$ is $1$.  
It is also easy to see that there are
at least two mutually orthogonal vectors $\ket{b}$ 
satisfying \vone .  For example, let $\ket{v_1} , \ket{v_2} ,
\ket{v_3}$ 
be orthonormal
vectors and take $\ket{a} = \ket{v_1}$ 
and $\ket{c} = \lambda \ket{v_1} + \mu \ket{v_2} $, where 
$\lambda^2 + \mu^2 = 1$ and we may take $\lambda$ and $\mu$ real.
Then the vectors 
\eqn\vecs{\ket{b_{\pm}} = \lambda \ket{v_1} + {{\mu} \over {x}} \ket{v_2} \pm 
{{ (x-1)^{1/2} \mu} \over {x}} \ket{v_3} }
both satisfy \vone\ and are orthogonal if $x$ is real and
$x^2 \lambda^2 = (x-2) ( 1 - \lambda^2 )$, which has solutions for 
$\lambda \leq 1/3$. 
Thus this construction produces 
consistent sets which give contrary 
probability one retrodictions. 
\medskip\noindent
{\it Example 2}  \qquad  Now consider the formalism applied to
generalised quantum mechanics, choose vectors $\ket{a}$, $\ket{b}$ 
and $\ket{c}$ as above,
take $\rho_i =  \ket{a} \bra{a}$ and 
$\rho_f = P_c / C$, 
where the normalisation constant 
$C =  \, | \braket{a}{c} |^2$.   
Suppose that the data correspond to the history 
$\H = \{ P_a \}$ from 
the consistent set $\S = \{  \{P_a , 1 - P_a \} \}$, 
and consider consistent extensions of the form
$\S' = \{ \{P_a , 1 - P_a \}, \{P_b , 1 - P_b \} \}$ where $P_b$ is as in 
Example 1.  As above, the conditional probability of $P_b$
given $H$ is $1$, so that we obtain consistent sets 
which give mutually contrary probability one predictions. 
\medskip\noindent
Note that it is impossible to produce an example 
in which the formalism makes contrary predictions when applied
to ordinary quantum mechanics.  In this context, if $P$ is predicted
with probability $1$ from the history $H = \{ P_1 , \ldots , P_n \}$ in 
the set $\S$, then 
\eqn\id{P P_n \ldots P_1 \rho^{1/2} = P_n \ldots P_1 \rho^{1/2} \, ,}
and 
\eqn\zer{Q P_n \ldots P_1 \rho^{1/2} = 0}
if $Q \leq (1-P)$. 
Thus if a projection $Q$ orthogonal to $P$ belongs to any
consistent set then its probability in that set, conditional on the 
history $H$, is zero. 
It is also easy to see that
we can construct examples in which any number of consistent sets
make mutually contrary retrodictions --- or, in the 
case of time-neutral quantum mechanics, predictions --- by 
taking the dimension of 
$\cal{H}$ to be sufficiently large and choosing $| \braket{a}{c} |$ 
sufficiently small.

Though the contrary inferences in the above example both
correspond to one-dimensional projections,
it is easy to construct similar examples of
contrary inferences corresponding to projections of different
dimension since, given the above initial and
final states, the condition for a projection $P$ to correspond
to a probability one consistent inference is simply that 
$\bra{a} P \ket{c} = \braket{a}{c} $. 
Thus contrary inferences could not be avoided by 
introducing a unitary equivalence relation --- perhaps along the
lines of those recently considered by Gell-Mann and 
Hartle\refs{\gmhequiv} --- according to which any pairs
of projections involved in contrary inferences are 
declared physically equivalent. 

\newsec{Conclusions} 

The incompatibility of the 
logics corresponding to different consistent sets is generally
described as a natural generalisation of the principle of 
complementarity in Copenhagen quantum mechanics: a discussion
making precisely this point can be found, for example, in
Chapter 5.4 of Omn\`es' recent book\refs{\omnesbook}. 
There is, though, no parallel in standard quantum mechanics 
for the prediction and retrodiction of contrary propositions, 
and many might feel that no acceptable interpretation of 
quantum theory should allow such inferences. 
Indeed, Omn\`es comments that ``The worst event would be if
two different ways of reasoning could lead to different conclusions
when one is using two different consistent logics.  In view of this
danger, which would mean that the present approach is completely 
wrong, we shall initially discuss how two different logics can be
related to each other.''\refs{\omnesquote}

Now Omn\`es has in mind here a slightly
different possibility, namely that if two propositions both belong to 
two distinct consistent sets, and one implies the other in one set, 
the implication might fail in the other set.  
This cannot happen in the consistent histories formalism.  
It is not possible, for
example, to use the same set of data to predict the proposition $P$ 
in one set and its negation $(1-P)$ in another, both with probability
one.  At first sight it may seem as though the above examples do 
precisely this.  The reason why they fail to do so is that, in the
consistent histories formalism, if we have two propositions
corresponding to projection operators $P \leq Q$ (i.e. the range
of $P$ is a subspace of that of $Q$) and if $P$ is predicted
with probability one, it does not follow that $Q$ is predicted 
with probability one (or with any other probability).  

It might possibly be argued that this last feature is 
less of a flaw, and that the examples above are 
less worrying, than the type of contradictory inference
Omn\`es considers --- but it is hard to see why.  
The fact that the theory stipulates that the pictures
corresponding to different sets are incompatible alternatives cannot 
be used as a defence here without allowing the same defence in the 
case of Omn\`es' hypothetical disaster.  
Clearly, no logical contradiction arises if we suppose that different
consistent sets simply give different pictures of the physics 
and if we make no stipulation whatsoever about the relationship 
between these pictures.  
But the same of course is true of the analogous
supposition about {\it inconsistent} sets of histories.
To justify the fundamental assumption of the consistent histories
formalism --- that it is precisely the 
consistent sets which give sensible
physical descriptions --- we need to suppose the following.
First, that it is wrong --- a product of misguided classical 
intuition --- to suppose that 
contrary propositions $P$ and $Q$ should never be inferred with
probability one in different sets.  Second, that it is right ---
a fundamental feature of quantum physics --- to suppose that 
contradictory propositions $P$ and $(1-P)$ should never be
inferred in different sets.  This is certainly not the standard
understanding of the situation, and there seems no obvious
reason to adopt it.  One might plausibly try to argue that 
both suppositions are right, or conceivably
(depending on how the descriptions are to be used) that 
both are wrong, but it seems particularly hard to argue 
for one and against the other.  

Another possible counter-argument is that, in the end, scientists 
need only worry about predictions, and contrary predictions can 
be avoided by restricting the formalism to standard, rather than 
generalised, quantum mechanics.   
One difficulty with this line of defence is that 
it is the retrodictive cosmological applications of 
the consistent histories 
formalism that are presently the most interesting. 
Unlike other approaches to quantum theory, the formalism allows us 
to discuss series of past cosmological events and to 
assign probabilities to them, even when some or all of the events
occurred before the formation of classical structures.  
Prediction, on the other hand,
is where the consistent histories formalism is at its weakest. 
No coherent interpretation of the formalism has been
found which unambiguously implies the standard predictions of Copenhagen
quantum mechanics, although those
predictions (among many others) can be reproduced
by calculations within the formalism. 
Moreover, though the formalism allows many different predictive
calculations, those which are new seem to be physically 
irrelevant except in highly implausible scenarios and, possibly,
in the case of generalised time-neutral quantum cosmology.  
This, though, is precisely the case in which contrary 
predictions arise.  

If we reject these defences we seem to be left with the
conclusion that the contrary inferences implied by the consistent histories
formalism make it hard to take it seriously 
as a fundamental theory in its present 
form.  This means that further constraints beyond 
consistency are needed in order to construct a natural
generalisation of the Copenhagen interpretation to closed systems. 
Whether physically
sensible and mathematically precise constraints can be found
in standard versions of the formalism, such as the one
above, is an important and intriguing open question.  
It might also be interesting to 
investigate the analogous problem in the more abstract schemes
characterising the logical structure of consistent histories 
which have recently been developed.\refs{\ishamlinden} 
\vfill\eject 
\noindent{\it Note added}\medskip

After writing this Letter, my attention was drawn to Cohen's 
interesting analysis\refs{\cohen} of an example due to Aharonov and 
Vaidman\refs{\av}.  Ref. \refs{\cohen} includes what seems to be the first 
consistent histories analysis of an example in which contrary 
inferences arise.\foot{The argument of 
the relevant section, VIII B, of Ref. \refs{\cohen} is not 
entirely correct: the statement that failure of consistency 
follows from the failure of the relevant projection operators to commute
is false.} 
Ref. \refs{\cohen} is a critique of Aharonov and Vaidman's arguments 
rather than those of the consistent
histories literature, so that the implications for the consistent 
histories formalism are not considered in any detail. 

A response to the present Letter, defending the 
consistent histories formalism, can be found in
Ref. \refs{\griffchqr}. 
Griffiths stresses the point, emphasized above, that the 
consistent histories formalism can be interpreted in a way which 
leads to no logical contradiction.  

As mentioned above, it seems that further natural constraints beyond
consistency seem to be needed for a sensible  
formulation of the quantum theory of closed systems. 
It turns out that at least one such constraint exists: 
a stronger version of the consistent histories formalism, designed
to avoid the problems discussed in this paper, is described
in Ref. \refs{\akordered}.  
\vskip15pt
\leftline{\bf Acknowledgements}  

I am very grateful to Oliver
Cohen, Fay Dowker, Arthur Fine and Lucien Hardy for helpful 
discussions and to Bob Griffiths and Jim Hartle for 
critical readings of the manuscript and valuable comments. 
I thank the Royal Society for financial support. 
\listrefs
\end